\documentclass{article}
\usepackage{spconf,amsmath,graphicx}
\usepackage{amsthm,amsmath,amssymb}
\usepackage{mathrsfs}
\usepackage{subfigure}
\usepackage{multirow}
\usepackage{bbding}
\usepackage{stfloats}
\usepackage{pifont}
\usepackage{booktabs}
\usepackage{hyperref}

\ninept 
\hypersetup{
    colorlinks=true,
    linkcolor=blue,
    filecolor=magenta,      
    urlcolor=cyan,
    pdftitle={Overleaf Example},
    pdfpagemode=FullScreen,
    }

\newcommand\blfootnote[1]{%
  \begingroup
  \renewcommand\thefootnote{}\footnote{#1}%
  \addtocounter{footnote}{-1}%
  \endgroup
}
\newcommand*{\email}[1]{%
    \normalsize\href{mailto:#1}{#1}\par
    }

\title{Improving Few-Shot Learning for Talking Face System with TTS Data Augmentation}
\name{Qi Chen\textsuperscript{1}, Ziyang Ma\textsuperscript{1}, Tao Liu\textsuperscript{1}, Xu Tan\textsuperscript{2}, Qu Lu\textsuperscript{3}, $^*$ Xie Chen\textsuperscript{1}, $^*$ Kai Yu\textsuperscript{1}}
\address{
\textsuperscript{1}X-LANCE Lab, Department of Computer Science and Engineering,\\MoE Key Lab of Artificial Intelligence, AI Institute, \\
Shanghai Jiao Tong University, China \\
\textsuperscript{2}Microsoft Research Asia,  \textsuperscript{3}Shanghai Media Tech \\
\email{\{cq1073554383,zym.22,liutaw,chenxie95,kai.yu\}@sjtu.edu.cn}}

\begin{document}
\maketitle

\begin{abstract}
Audio-driven talking face has attracted broad interest from academia and industry recently. However, data acquisition and labeling in audio-driven talking face are labor-intensive and costly. The lack of data resource results in poor synthesis effect. To alleviate this issue, we propose to use TTS (Text-To-Speech) for data augmentation to improve few-shot ability of the talking face system. The misalignment problem brought by the TTS audio is solved with the introduction of soft-DTW, which is first adopted in the talking face task. Moreover, features extracted by HuBERT are explored to utilize underlying information of audio, and found to be superior over other features. The proposed method achieves 17\%, 14\%, 38\% dominance on MSE score, DTW score and user study preference repectively over the baseline model, which shows the effectiveness of improving few-shot learning for talking face system with TTS augmentation.

\blfootnote{$^*$ Xie Chen and Kai Yu are the
corresponding authors.}
\blfootnote{$^\dag$ Code and demo are available at \url{https://github.com/Moon0316/T2A}}

\end{abstract}

\begin{keywords}
few-shot, soft-DTW, HuBERT, low-resource
\end{keywords}

\section{Introduction}
\label{sec:intro}
Recent years have witnessed digital human becoming a popular topic in academia and industry. It has numerous applications in the media and entertainment industry, including virtual news reporters, virtual YouTubers, virtual idols, and so on. In order to generate a high-quality digital human, it is crucial to synthesize natural face motions, i.e., talking face generation. There are two main categories of talking face, video-driven and audio-driven. Video-driven talking face, also known as facial reenactment, focuses on driving a portrait with a reference video \cite{thies2016face2face,zollhofer2018state,nagano2018pagan}. However, video-driven talking face sometimes suffers from the unstable reenactment of face motions. Audio-driven talking face aims to synthesize face motions corresponding to the given audio and is more stable in face motion synthesis \cite{NVP, S2A, makeittalk}.


Face motion representation in audio-driven talking face is a key factor at the system level. 
Commonly used face motion representations in the pipeline include landmark-based formats like face mesh \cite{voca,faceformer} or motion-based formats like blendshape \cite{S2A}. Acquiring these types of corpus is challenging, as they are recorded by professional devices, which hinders their wide application. Although some mobile devices can record such type of data in a convenient way, the collected data is not accurate because of the limited accuracy of devices. As a result, data need to be artificially refined before training, which consumes high labor costs. Training with few-shot learning remains to be investigated.



\begin{figure}[t]
    \centering
    \includegraphics[width=0.95\linewidth]{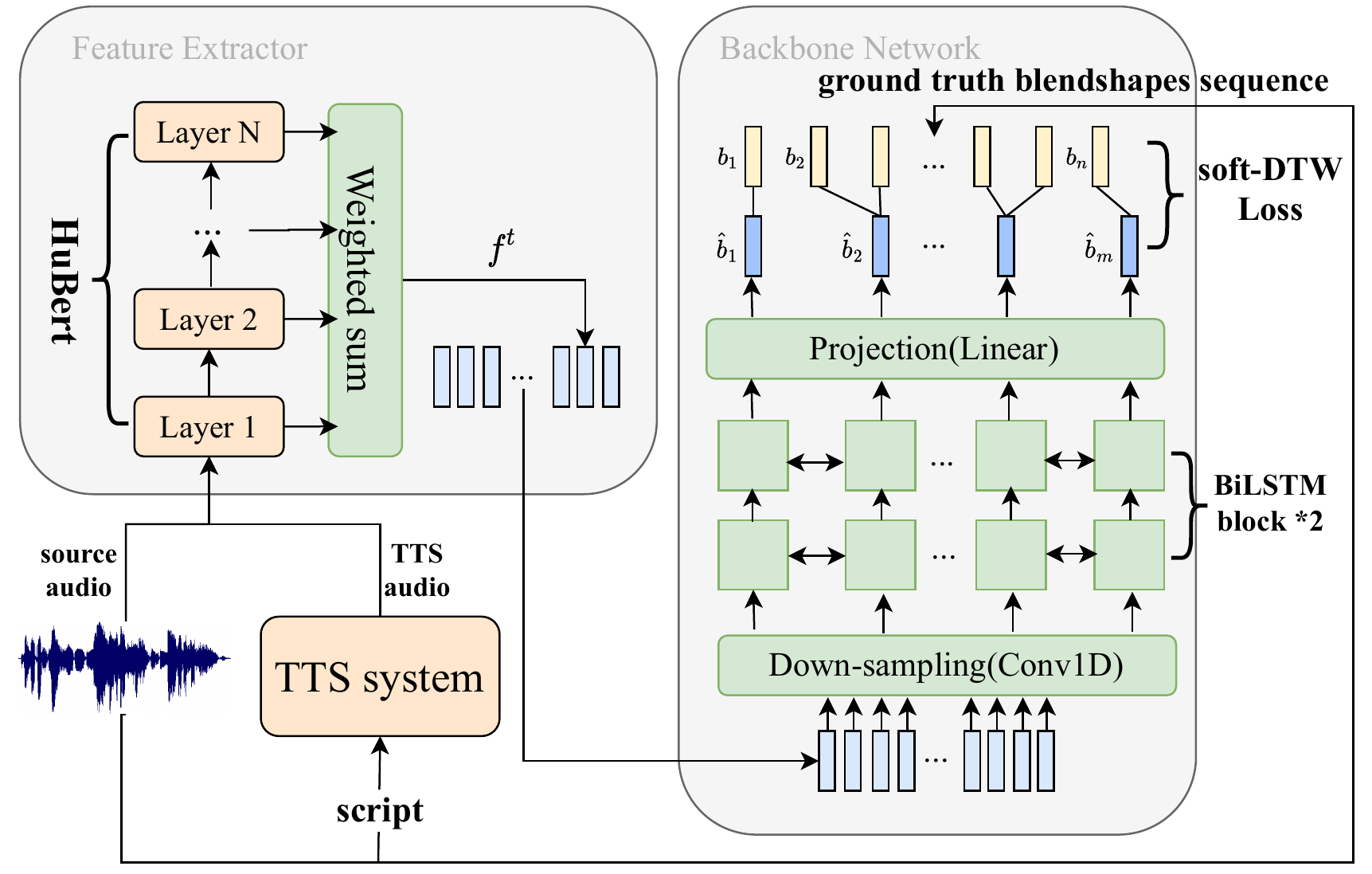}
    \caption{The overall architecture of the proposed TTS-augmented audio-driven talking face.} 
    \label{fig:arch}
    \vspace{-0.7cm}
\end{figure}

With the development of text-to-speech (TTS) \cite{ren2020fastspeech,du2021rich,du2022vqtts}, off-the-shelf TTS systems are able to synthesize high-quality audio, and are frequently utilized for data augmentation to improve speaker diversity in various speech-related tasks, such as speech recognition \cite{du2020speaker} and speaker verification \cite{du2021synaug}. Besides, pre-trained models (PTMs) trained on large amounts of unlabeled audio data shows promising potential for audio feature extraction. 
Therefore, combining TTS data augmentation and PTMs is a feasible method to boost low-resource talking face generation. 



In this paper, to alleviate the issues above and improve the robustness of the talking face, we adopt a TTS augmentation strategy to produce extra training data and combine them with a few labeled data. To address the misalignment issue that existed in the newly-generated data, we introduce soft-DTW \cite{cuturi2017soft}, a differentiable formulation of dynamic time warping (DTW) \cite{dtw4} algorithm, to align the training data. To the best of our knowledge, this loss function is first introduced to the talking face task. Then, we investigate a robust audio feature utilizing HuBERT \cite{hsu2021hubert} model. The weighted sum of features is adopted to fully utilize the underlying information of audio. Experiments show that our method achieves significant improvements on a small training corpus of ten randomly picked utterances for about one minute. Code and demo are released, and we hope our method will contribute to the community.

\section{Related works}
\label{sec:format}
\subsection{Audio-driven Talking Face}
Audio-driven talking face aims to generate face motions aligned with the input speech signal. There are two types of audio-driven talking face systems in the literature. One is the modularized system, and the other is the end-to-end system. Although there are active research efforts and promising research progress reported in the end-to-end talking face system \cite{wav2lip}, the modularized is still the mainstream system used in practice so far, which is also the main focus of this paper. A typical pipeline of the modularized system consists of two modules,
first mapping the acoustic features to some face motion representations and then rendering video from the predicted face motion representation. 
There are various options for the acoustic features and face motion representations in previous works. Thies et al.  \cite{NVP} utilized DeepSpeech \cite{deepspeech} to extract the audio feature, and a CNN-based model was followed to map the audio feature to the 3D Morphable Model (3DMM) features, which was then used to generate the neural texture, finally image2image translation method was adopted to render photo-realistic videos. Chen et al.  \cite{S2A} used phonetic posteriorgrams (PPGs) along with pitch and energy as audio features, which were mapped to blendshapes with a mixture-of-experts (MOE) Transformer, and finally rendered into animation by a rendering engine. Zhou et al. \cite{makeittalk} disentangled linguistic information and speaker information of the audio by a voice conversion model AutoVC \cite{autovc} and applied the LSTM network to predict facial landmarks, and finally synthesize photo-realistic videos by image2image translation or synthesize animation by face warping.

\subsection{HuBERT}
With the explosive growth of the amount of raw audio data on the Internet, the effective utilization of these unlabeled data has become a research hotspot. 
Pre-trained models (PTMs) using self-supervised learning have achieved tremendous success in a range of speech-related tasks~\cite{schneider2019wav2vec, baevski2019vq, baevski2020wav2vec, hsu2021hubert,baevski2022data2vec, ma2022mt4ssl}. Among them, HuBERT is one of the best-performing models in the field of speech representation learning.
In the pre-training stage, HuBERT is trained by predicting discrete target labels with a BERT-like prediction loss. Discrete labels are obtained by performing K-means clustering on raw audio itself without the use of any labeled data. 
In the fine-tuning phase, supervised learning is used to adapt the model to a specific downstream task, such as automatic speech recognition (ASR).
The recent SUPERB benchmark~ \cite{yang2021superb} demonstrates that the features extracted from self-supervised pre-training models exhibit universal superior performance over traditional handcraft acoustic features such as MFCC on various downstream tasks. 

\subsection{Dynamic Time Warping}
Dynamic Time Warping (DTW) \cite{dtw4} is a dynamic programming algorithm that decides the best alignment of two sequential data and quantifies the degree of similarity. DTW is able to calculate the minimum distance of two sequences with different lengths by warping the time axis of the series and performing a dynamic programming algorithm. It was originally designed to tackle the problem of speaking rate variation under the same speaking pattern in spoken word recognition tasks \cite{dtw4}, and successfully quantified the similarity of the length variable speech sequence with the word template. Donald et al. \cite{dtw5} used DTW in pattern-matching problems and found that DTW shows excellent potential in finding patterns in time series.

\section{Methods}
\label{sec:Methods}
\subsection{Model Architecture for TTS-augmented system}
As shown in Figure \ref{fig:arch}, the source audio and TTS audio are fed to the feature extractor module for acoustic feature extraction, followed by a lightweight BiLSTM-based backbone network for face motions prediction. 
We use blendshape format as the face motions representation in this work. 
Finally, UE4 (Unreal Engine 4) is adopted to render the predicted blendshapes sequence into animation \footnote{The discussion of rendering is out of scope and is not detailed here.}. 


\textbf{Augmented Audio-animation Pairs Generation.} 
\quad The augmented audio-animation pair is generated by converting the script of the source audio to the TTS audio with an off-the-shelf TTS system and taking the blendshapes matching with the source audio as the label of the TTS audio. However, the TTS model has its own duration predictor, which results in the duration variation between the source audio and the TTS audio at the word and the sequence level, and brings difficulty in aligning the TTS audio and the reference blendshapes. As a result, the reference blendshape of each acoustic feature frame is not determinate as that in the source audio-animation data. The problem of duration misalignment in the TTS audio-animation pair will be solved in \ref{sec:softdtw}.

\textbf{Feature Extractor.} \quad To better utilize the underlying structure information lied in the audio, a pre-trained HuBERT\cite{hsu2021hubert,pretrained} model is adopted as feature extractor, and the hidden layer outputs of each layer are weighted and summed to obtain the acoustic feature as the input for the talking face system, which can be formulated as:

$$
\setlength{\abovedisplayskip}{2pt}
\setlength{\belowdisplayskip}{2pt}
f^t = \sum_{i=1}^{N}(\alpha_i*h_i^t) \quad , \quad
\sum_{i=1}^{N}\alpha_i =1
$$
Where $h_i^t$ denotes the i-th layer output of HuBERT at time t, $N$ is the total number of hidden layers in HuBERT, $\alpha_i$ represents the learnable weight of $h_i^t$, and $f^t$ is the final acoustic feature at time t.

\textbf{Backbone Network.} 
\quad We first adopt a convolutional module to downsample the features extracted by HuBERT from 50Hz to 25Hz aligned with blendshapes. The convolutional downsampling module is a 2-layer 1-D convolutional neural network.
The downsampled features are then fed to a 2-layer lightweight BiLSTM network, followed by a linear layer for the blendshapes sequence prediction.


\subsection{Soft-DTW Loss for TTS Audio Alignment}
\label{sec:softdtw}
DTW can measure the distance of two sequences with different lengths, which fits the case in our task as the TTS audio and the ground truth blendshapes sequence are not guaranteed to have the same length. However, DTW is nondifferentiable because of the min operation in the dynamic programming process. In order to derive a differentiable form of DTW computation, soft-DTW \cite{cuturi2017soft} is adopted. The soft-DTW loss allows soft alignment between the predicted blendshapes sequence derived from audio and the ground truth blendshapes sequence, and so as to enable the use of TTS data with variable lengths. To the best of our knowledge, this loss function is first introduced to the talking face task.

To formulate the soft-DTW loss, we first consider the DTW distance of the predicted blendshapes sequence $(\hat{b}_1,\hat{b}_2,\cdots,\hat{b}_m)$ and the ground truth blendshapes sequence $(b_1,b_2,\cdots,b_n)$. Through the DTW computation, tokens from two sequences will be aligned in pairs, and a feasible path can be defined as an ordered sequence of token pairs that submits to some restrictions, the target of DTW is to find a feasible path with minimum cost. To acquire the formulation of DTW, denoting the $k$-th paired item in a specific path as $(\hat{b}_{i_k},b_{j_k})$, two restrictions in DTW are exerted as follows: 
\begin{equation}
\setlength{\abovedisplayskip}{2pt}
\setlength{\belowdisplayskip}{2pt}
\begin{aligned}
i_{k+1}-i_k\leq1, \quad j_{k+1}-j_k\leq1 & \quad\cdots\cdots  continuity \\
i_{k+1}-i_k+j_{k+1}-j_k\geq1  & \quad\cdots\cdots monotonicity
\label{eq:res}
\end{aligned}
\end{equation}
Denoting $\delta_{ij}$ as the distance of token $\hat{b}_i$ and $b_j$. In our work, Euclidean distance is adopted to measure the distance of two tokens, so $\delta_{ij}=||\hat{b}_i-b_j||_2$. The cost of a feasible path is the sum of the distances of all paired tokens in the path. To avoid the enumeration of all feasible paths, dynamic programming is adopted to find the optimal path. Denoting the cost of the optimal DTW path between sequence $(\hat{b}_1,\cdots,\hat{b}_i)$ and sequence $(b_1,\cdots,b_j)$ as $r_{i,j}$, and our target is equivalent to find the minimum $r_{m,n}$. According to Eq.\ref{eq:res}, state transition equation of the optimal path cost is formulated as:
\begin{equation}
\setlength{\abovedisplayskip}{2pt}
\setlength{\belowdisplayskip}{2pt}
    r_{i,j} = \delta_{ij}+\min\{r_{i-1,j},r_{i,j-1},r_{i-1,j-1}\}
    \label{eq:trans}
\end{equation}

The min operator prevents the differential operation of $r_{i,j}$. To solve this problem, soft-DTW introduces a smoothed formulation of the min operator, that is:
\begin{equation}
\setlength{\abovedisplayskip}{2pt}
\setlength{\belowdisplayskip}{2pt}
\min^\gamma\left\{d_1, \ldots, d_n\right\}:= \begin{cases}\min _{i \leq n} d_i, & \gamma=0 \\ -\gamma \log \sum_{i=1}^n e^{-d_i / \gamma}, & \gamma>0\end{cases}
\label{eq:min}
\end{equation}

When $\gamma>0$, $\min^\gamma\left\{d_1, \ldots, d_n\right\}$ is differentiable, and the closer $\gamma$ is to $0$, the closer $\min^\gamma\left\{d_1, \ldots, d_n\right\}$ is to $\min\left\{d_1, \ldots, d_n\right\}$. Soft formulation of Eq. \ref{eq:trans} can be derived from Eq. \ref{eq:min}, where $r_{i,j}$ is differentiable:
\begin{equation}
\setlength{\abovedisplayskip}{2pt}
\setlength{\belowdisplayskip}{2pt}
    r_{i,j} = \delta_{ij}-\gamma \log (e^{-r_{i-1,j}/\gamma}+e^{-r_{i,j-1}/\gamma}+e^{-r_{i-1,j-1}/\gamma})
    \label{eq:softtrans}
\end{equation}
Finally, the soft-DTW loss is derived as follows, the calculation of $r_{m,n}$ follows the state transition equation Eq. \ref{eq:softtrans}, which requires a time complexity of $O(mn)$.
\begin{equation}
\setlength{\abovedisplayskip}{2pt}
\setlength{\belowdisplayskip}{2pt}
    \mathcal{L}_{soft-DTW} = r_{m,n}
    \label{eq:softdtw}
\end{equation}
According to the chain rule, $\frac{\partial \mathcal{L}}{\partial \delta_{ij}}$ can be calculated from Eq. \ref{eq:softtrans}, please refer to \cite{cuturi2017soft} for more mathematical details of gradient calculation. With the back propagation of $\frac{\partial \mathcal{L}}{\partial \delta_{ij}}$, model parameters will be updated during the training process.

\vspace{-0.3cm}
\section{Experiments}
\label{sec:exp}
\vspace{-0.15cm}
\subsection{Experiment Settings}
\vspace{-0.4cm}
\begin{table}[!htp]
    \centering
    \caption{Speakers allocation of datasets}
    \setlength{\tabcolsep}{1mm}{\begin{tabular}{c c c c}
    \toprule
    \textbf{config name}    & \textbf{train} & \textbf{validation} & \textbf{test} \\
    \midrule
      REC   & record & record & \multirow{4}{*}{record} \\
      TTS1   & 1 TTS speaker & 1 TTS speaker &  \\
      TTS13   & 13 TTS speakers & 13 TTS speakers &  \\
      REC+TTS13   & record+13 TTS speakers & record &  \\
    \bottomrule
    \end{tabular}}
    \label{tab:dataset}
\end{table}

\textbf{Dataset Description} \quad We use a Chinese dataset from the 2022 text\&audio-driven talking face competition of World Artificial Intelligence Innovation  (AIWIN)\footnote{\url{http://ailab.aiwin.org.cn/competitions/69}}. The dataset contains three splits, including train (276 sequences, 29m8s in duration), validation (79 sequences, 9m58s in duration), and test set (85 sequences, 10m30s in duration). There is a single female speaker in this dataset (denote the audios of this speaker as ``record", the scripts of audios are available), and the neural TTS system from Microsoft Azure is used for TTS audio generation.
Four different dataset configurations are used for the experiment, they are REC, TTS1, TTS13, and REC+TTS13 in Table \ref{tab:dataset}, where ``record" are recorded audio synchronized with the ground truth animation, while TTS audios are variable in duration and are not aligned with the ground truth animation. All dataset configurations take the audio-animation data from the recorded audio as the test set for a fair comparison. 

\textbf{Experiment Setups} \quad For each experiment, we trained the model for 30 epochs and selected the best model by evaluating soft-DTW loss on the validation set. Finally, we render the output animations from the predicted blendshapes to videos by Unreal Engine 4 (UE4). The demo can be found on our project page.

\textbf{Evaluation Metrics} \quad To analyze experiment results, we performed objective and subjective evaluations. During the objective evaluation, the MSE and DTW scores of models are tested on the ``record" test set. The audios and animations in the ``record" test set are strictly aligned, and the MSE score is able to average the distance between each aligned token pair, so MSE score quantifies the similarity of sequences at the frame level. As previous research \cite{dtw5} demonstrated DTW's capability of finding patterns in time series, the DTW score is introduced additionally and is expected to reflect the overall performance at the sequence level. For the subjective evaluation, a user study in the form of AB tests is conducted. In the AB tests, we presented volunteers with animations predicted by different models and asked them to choose the animation that synchronizes with the background audio better. The audios used in the AB tests consist of 2 Chinese News report audios, 1 Chinese ``record" audio, 1 Chinese TTS audio, and 1 English audio. We collected 41 pieces of feedback in total to calculate the preference rate for each model.

\vspace{-0.1cm}
\subsection{TTS Data Augmentation}
In the following experiment, the effectiveness of TTS data augmentation on the low-resource scenario will be explored. The baseline model is trained on ten sequences (70s in duration) from the REC dataset. In contrast, TTS audios from 13 other speakers are generated based on the transcription of 10 sequences, and are combined with the recorded audios to enhance the baseline talking face system with the soft-DTW algorithm. The experiment results can be found in Table \ref{tab:dataaug}.


\begin{table}[!hpt]
  \caption{Objective and subjective results of models trained by recorded audios, recorded audios + TTS audios of 13 speakers on merely 70 seconds labeled data}
  \centering
  \begin{tabular}{ccc}
    \toprule
     \textbf{ Metrics } & \textbf{REC} & \textbf{REC+TTS13} \\ 
    \midrule
 MSE score       & 0.00473 & \textbf{0.00391} \\
 DTW score       & 0.68819 & \textbf{0.54433} \\
 \midrule
 User study preference rate & 31.0\% & \textbf{69.0\%} \\
    \bottomrule
  \end{tabular}
    \label{tab:dataaug}
\end{table}

According to the results, the TTS augmented system yields consistent and noticeable performance improvement over the baseline in objective and subjective evaluations. The rendering results of our system are also satisfactory, which demonstrate that our proposed method can build a decent audio-driven talking face system with merely 70 seconds of labeled audio-animation data. For more information, readers are referred to the demo on the project page.

\vspace{-0.2cm}
\subsection{TTS-driven talking face}
The next experiment is conducted to compare the performance of the talking face system trained on TTS audios and that trained on recorded audios. To this end, three systems were trained based on REC, TTS1, and TTS13 datasets as defined in Table \ref{tab:dataset}. The experimental results are shown in Table \ref{tab:t2a}.

\vspace{-0.2cm}
\begin{table}[!hpt]
  \caption{Objective and subjective results of talking face models trained on recorded audios, TTS audios with one speaker, TTS audios with 13 speakers on 29m8s training data}
  \centering
  \begin{tabular}{cccc}
    \toprule
     \textbf{ Metrics } & \textbf{REC} & \textbf{ TTS1} & \textbf{TTS13} \\ 
    \midrule
 MSE score & \textbf{0.00349} & 0.00420 & 0.00378  \\
 DTW score & \textbf{0.46135} & 0.53501 & 0.49961  \\
 \midrule
 User study preference rate & \textbf{61.0\%} & - & 39.0\% \\
    \bottomrule
  \end{tabular}
    \label{tab:t2a}
\end{table}

According to the objective and subjective results, the talking face model trained on the TTS data is slightly behind that trained on recorded audio, which is expected as the latter adopted the synchronized audio, and the model is evaluated in the matched recorded audio. 
Despite this gap, the performance on TTS13 is still impressive, as there is zero synchronized recorded speech used during training. It is also noteworthy that the TTS13 system outperforms the TTS1 system, which indicates that the increase of diversity in the train set helps to improve performance. The result that TTS-driven talking face system is able to achieve satisfactory performance demonstrates the effectiveness of combining TTS audio with soft-DTW loss in the audio-driven talking face from another aspect.

\vspace{-0.2cm}
\subsection{Model Robustness}
\begin{figure}[htbp]
	\centering
	\includegraphics[width=0.7\linewidth]{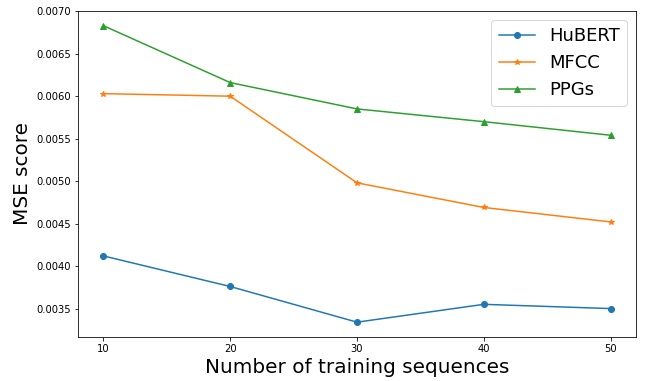}
	\caption{MSE score of models trained by HuBET, MFCC, PPGs features in data resource of 10, 20, 30, 40, 50 training sequences}
	\label{fig:feature_contrast}

\end{figure}

\textbf{Ablation study on Feature Robustness.} \quad Weighted sum of HuBERT features is adopted in the proposed method. To validate the robustness and superiority of the HuBERT feature, we compared the talking face models with different types of audio features in various amounts of labeled data. 
The audio features include HuBERT feature,  MFCC, phonetic posteriorgrams (PPGs), where PPGs are the last layer output of an ASR model trained on the Chinese speaking corpus AISHELL-2 \cite{du2018aishell}. The results are shown as Figure \ref{fig:feature_contrast}. It turns out that our HuBERT feature significantly and consistently outperforms the other two types of features and yields the best performance. More details can be found on the project page.

\begin{figure}[!htbp]
	\centering
	\setlength{\abovedisplayskip}{0pt}
	\setlength{\belowdisplayskip}{0pt}
	\includegraphics[width=0.7\linewidth]{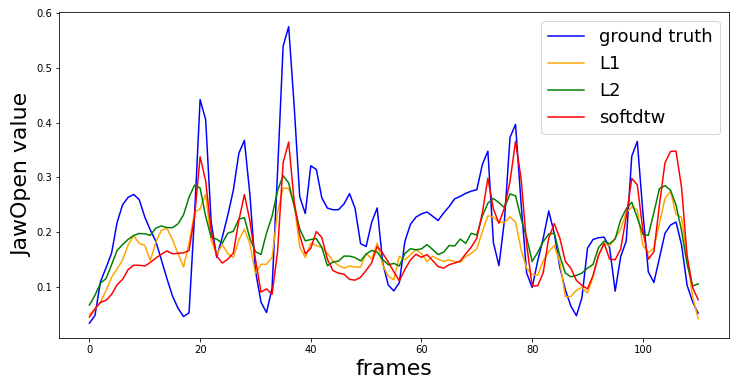}
	\caption{JawOpen value change trend of ground truth and value predicted by models trained with L1, L2, soft-DTW losses}
	\label{fig:jawopen}
\end{figure}

\textbf{Ablation study on Loss Function Effectiveness.} \quad The next experiment is to demonstrate the effectiveness of the soft-DTW loss in comparison with the conventional L1 and L2 losses, which are commonly used on talking face tasks. During the experiments, the dataset configuration is fixed as REC. Objective evaluation results for models trained by different loss functions are shown in Table \ref{tab:loss}. We can find that models trained by L1, L2, and soft-DTW yield similar MSE scores, which indicates that the soft-DTW loss has comparable fitting performance with L1 and L2
losses at the frame level. Meanwhile, the model trained by soft-DTW exceeds L1 and L2 losses largely on the DTW score, indicating it has the potential to better predict sequences with similar shapes as the ground truth. To compare the ability of fitting sequence shape, we randomly pick a piece of audio in the test set and plot the ``JawOpen'' value(a dimension of blendshapes). Figure \ref{fig:jawopen} shows that model trained by soft-DTW is better at predicting the peaks in a blendshapes sequence, which is expected to result in a more expressive rendered animation. 

\vspace{-0.2cm}
\begin{table}[!hpt]
  \caption{Performance of models trained by L1, L2, soft-DTW losses}
  \centering
  \begin{tabular}{cccc}
    \toprule
     \textbf{ Metrics } & \textbf{ L1} & \textbf{L2} & \textbf{soft-DTW} \\ 
    \midrule
 MSE score       & \textbf{0.00344} & 0.00351 & 0.00349\\
 DTW score       & 0.51082 & 0.55999 & \textbf{0.46135} \\
    \bottomrule
  \end{tabular}
    \label{tab:loss}
\end{table}
\vspace{-0.4cm}

\textbf{Ablation study on Model Architecture.} \quad To further verify the effectiveness of our method, we also changed the backbone network from BiLSTM to GRU, RNN and Transformer, and respectively trained models with or without TTS data augmentation. The results are shown as Table \ref{tab:modelarch}, TTS data augmentation methods show superiority on all of the 4 architectures. 


\begin{table}[!hpt]
  \caption{MSE score of models with different network architectures trained by recorded audio, recorded audio + TTS audio of 13 speakers on merely 70 seconds labeled data}
  \centering
  \begin{tabular}{ccccc}
    \toprule
      & \textbf{BiLSTM} & \textbf{GRU} & \textbf{RNN} & \textbf{Transformer}\\ 
    \midrule
 w/o aug       & 0.00473 & 0.00438 & 0.00447 & 0.00896\\
 with aug      & \textbf{0.00391} & \textbf{0.00406} & \textbf{0.00409} & \textbf{0.00632}\\
    \bottomrule
  \end{tabular}
    \label{tab:modelarch}
\end{table}

\vspace{-0.3cm}
\section{Conclusion}
We have developed a TTS data augmentation method in talking face tasks by producing augmented audio-animation pairs with a TTS system, and solved the misalignment problem brought by TTS audio with the introduction of soft-DTW loss. The weighted sum of HuBERT features is adopted to fully utilize the underlying information of audio. From objective and subjective experiments, our proposed method is proven to boost the few-shot ability of a talking face system in low data resources. For future studies, we are going to apply our TTS data augmentation method in other cases of talking face generation, such as synthesizing photo-realistic talking face. Hopefully, our method will serve as an alternative for increasing speaker diversity and improving few-shot ability in talking face tasks.

\textbf{Acknowledgements: }This work was supported by National Social Science Foundation of China (No. 6220070337 and 18ZDA293), 
State Key Laboratory of Media Convergence Production Technology and Systems Project (No.SKLMCPTS 2020003), 
Shanghai Municipal Science, Technology Major Project (2021 SHZDZX0102) and the International Cooperation Project of PCL. 

\vfill\pagebreak

\bibliographystyle{IEEEbib}
\bibliography{strings,refs}

\end{document}